\documentclass{elsarticle}
\usepackage{graphicx}
\usepackage{amssymb}
\usepackage{amsmath}

\newcommand{\be}{\begin{eqnarray}}
\newcommand{\ee}{\end{eqnarray}}
\newcommand{\Pnl}{P^\mathrm{nl}}
\newcommand{\Plinj}{P^\mathrm{lin}_{j}}
\newcommand{\Plinjp}{P^\mathrm{lin}_{+,j}}
\newcommand{\Plinjm}{P^\mathrm{lin}_{-,j}}
\newcommand{\Plinjpm}{P^\mathrm{lin}_{\pm,j}}
\newcommand{\Plin}{P^\mathrm{lin}}

\newcommand{\hMpci}{h\,\mathrm{Mpc}^{-1}}
\newcommand{\Mpci}{\mathrm{Mpc}^{-1}}

\newcommand{\hiMpc}{{h^{-1}\mathrm{Mpc}}}
\newcommand{\hikpc}{{h^{-1}\mathrm{kpc}}}
\newcommand{\himsun}{{h^{-1}M_\odot}}
\newcommand{\Omegam}{{\Omega_\mathrm{m}}}

\newcommand{\Omegab}{{\Omega_\mathrm{b}}}
\newcommand{\As}{{A_\mathrm{s}}}
\newcommand{\ns}{{n_\mathrm{s}}}

\begin{document}

\begin{frontmatter}

\title{Response function of the large-scale structure of the universe to the small scale inhomogeneities} 

\author[SU,IPMU,CREST]{Takahiro Nishimichi\corref{cor1}}
\author[SU,CEA]{Francis Bernardeau}
\author[Y,IPMU]{Atsushi Taruya}
\cortext[cor1]{Corresponding author}
\address[SU]{Sorbonne Universit\'es UPMC Univ Paris 6 et CNRS, UMR 7095, Institut dÕAstrophysique de Paris, 98 bis bd Arago 75014 Paris, France}
\address[CEA]{CEA - CNRS, UMR 3681, Institut de Physique Th\'eorique, F-91191 Gif-sur-Yvette, France}
\address[IPMU]{Kavli Institute for the Physics and Mathematics of the Universe (WPI),
The University of Tokyo Institutes for Advanced Study,
The University of Tokyo, 5-1-5 Kashiwanoha, Kashiwa 277-8583, Japan}
\address[CREST]{CREST, JST, 4-1-8 Honcho, Kawaguchi, Saitama, 332-0012, Japan}
\address[Y]{Yukawa Institute for Theoretical Physics, Kyoto University, Kyoto 606-8502, Japan}


\begin{abstract}
In order to infer the impact of the small-scale physics to the large-scale properties of the universe, 
we use a series of cosmological $N$-body simulations of self-gravitating matter inhomogeneities
to measure, for the first time, the response function of such a system defined as a functional derivative
of the nonlinear power spectrum with respect to its linear counterpart.
Its measured shape and amplitude are found to be in good agreement 
with perturbation theory predictions except for the coupling from small to large-scale perturbations. 
The latter is found to be significantly damped, following a Lorentzian form. 
These results shed light on validity regime of perturbation theory calculations 
giving a useful guideline for regularization of small scale effects in analytical modeling.
Most importantly our result indicates that the statistical properties of the large-scale structure of the universe 
are remarkably insensitive to the details of the small-scale physics, astrophysical or gravitational, 
paving the way for the derivation of robust estimates of theoretical uncertainties on the determination of
cosmological parameters from large-scale survey observations.
\end{abstract}

\begin{keyword}
Gravitational growth of cosmic structures \sep Perturbation theory \sep $N$-body simulation
\end{keyword}

\end{frontmatter}

\section{introduction}
The cosmic energy fluctuations on large scales 
provide rich probes of the early universe physics, the mass of neutrinos 
or the nature of dark energy.
Wide-field galaxy surveys are therefore widely considered for unveiling the details 
of the universe \cite{Albrecht:2006lr}. Among them are the DES\footnote{https://www.darkenergysurvey.org/},  LSST\footnote{http://www.lsst.org/lsst/} and 
Euclid\footnote{http://www.euclid-ec.org} now under development. 
Such measurements rely however largely on our understanding of the statistical properties of the cosmic fluctuations. 
The great success of the latest cosmic microwave background observations
in establishing the standard picture of our Universe largely owed to the fact that the measured 
temperature fluctuations are in the linear regime and thus can accurately be predicted using linear theory
\cite{Hinshaw:2013uq,Planck-Collaboration:2015fj}.
Likewise, we expect that the late-time fluctuations on large scales are in a mildly nonlinear 
stage, and there are robust ways to predict them precisely beyond linear-theory calculations. 

Established probes such as the baryon acoustic oscillations (BAOs; e.g.~\cite{Eisenstein05, Cole:2005aa}) 
that give us a robust standard ruler useful for dark energy studies, or the redshift-space distortions 
(e.g.,~\cite{Kaiser87})
as an additional clue to discriminate gravity theories \cite{Linder08}, 
are among those that we expect in a mildly nonlinear regime. 
Alternatively, we can access cosmic fluctuations on similar and somewhat smaller scales 
with weak-lensing measurements (see \cite{Kilbinger:2014kx} for a recent review).
Such scientific programs can only be achieved if related observables
can be accurately predicted either from numerical simulations or analytically for any given cosmological model. 
In particular it is important that such observables are shielded from the details of astrophysics at galactic or sub-galactic scales 
\footnote{For instance in \cite{Jing:2006fk,Semboloni:2011qy,Mohammed:2014lr} baryonic effects are shown to be confined within the cluster scale, 
and they contribute to the matter power spectrum at most $\sim 10\%$ at $k=1\,\hMpci$ and drops rapidly toward larger scales.}.

One way to reformulate this question is to quantify the impact of small-scale structures on the growth of large scale modes. 
Perturbation theory (PT) is a powerful framework to predict the growth of structure.  
Assuming that the system is described by self-gravitating pressure-less fluids, 
it provides the first-principle approach to the nonlinear growth (see \cite{Bernardeau02} for a review). 
Its importance has been heightened after the detection of BAOs in the clustering of galaxies, 
making precise predictions of nonlinearities crucially important. 

PT calculations show precisely that mode couplings between different scales are unavoidable.
We propose here to quantify these couplings with a two-variable response function 
\footnote{This concept was recently utilized in Ref.~\cite{Taruya:2012qy} to compute the difference of the nonlinear power spectrum 
for slightly different cosmological models.}, defined as the linear
response of the \textit{nonlinear} power spectrum at wave mode $k$ with respect to the \textit{linear} counterpart
at wave mode $q$ \footnote{The normalization is such that $K$ contributes to the change in $\Pnl$ with uniform weights per decade.}:
\be
K(k, q; z) = q \frac{\delta \Pnl(k; z)}{\delta \Plin(q; z)}.
\label{eq:Kdef}
\ee
In the context of PT calculations, \cite{Bernardeau:2014lr,Blas14} showed progressive broadening of the response function 
with increasing PT order, pointing to the need of regularization of the small-scale contribution.

If the broadness of the response function at late times is true, physics at very small scale
can influence significantly the matter distribution on large scales, where the acoustic feature is prominent
\footnote{Notice, however, that the feature can also be affected by galaxy bias  
\cite{Angulo:2014lr,Prada:2014lr}.}.
It also questions the reliability of simulations, which can follow the evolution of Fourier modes only 
in a finite dynamical range.
We here discuss the response function at the non-perturbative level utilizing cosmological $N$-body simulations.

\section{Methodology}
We here describe our method to measure the response function from simulations.
We prepare two initial conditions with small modulations in the linear spectrum over a finite interval of 
wave mode $q$, evolve them to a late time, and take the difference of the nonlinear spectra 
measured from the two. That is
\be
\hat{K}_{i,j}\Plinj \equiv \frac{\Pnl_i[\Plinjp] - \Pnl_i[\Plinjm]}{\Delta \ln \Plin \Delta \ln q},
\label{eq:estimator}
\ee
where the two perturbed linear spectra are given by
\be
\ln\left[\frac{\Plinjpm(q)}{\Plin(q)}\right] = \left\{
\begin{array}{ll}
\displaystyle\pm\frac12\Delta\ln \Plin & \mathrm{if}\,\,q\in[q_j,q_{j+1}),\\
\displaystyle 0 & \mathrm{otherwise}.
\end{array}
\right.
\label{eq:plinjpm}
\ee
In the above, the index $i$ ($j$) runs over the wave-mode bins for the nonlinear (linear) spectrum, 
and we choose log-equal binning, $\ln q_{j+1} - \ln q_j = \ln k_{i+1} - \ln k_i = \Delta \ln q$. 
It is straightforward to show that the estimator $\hat{K}$ approaches to the response function $K$ defined in 
Eq.~(\ref{eq:Kdef}), when $\Delta \ln q$ and $\Delta\ln \Plin$ are small.
The definition~(\ref{eq:Kdef}) is advantageous in that it allows the measurement in this way at the fully nonlinear level
\footnote{This is contrasted to the function $F_n$ appearing in PT 
for the $n$-th order coupling.}.
Note that a similar function was first discussed numerically in Ref.~\cite{Neyrinck:2013lr} in the context 
of local transformations of the density field.

We adopt a flat-$\Lambda$CDM cosmology consistent with the five-year WMAP result \cite{WMAP5} with 
parameters $(\Omegam, \Omegab/\Omegam, h, \As, \ns) = (0.279, 0.165, 0.701, 2.49\times10^{-9}, 0.96)$,
which are the current matter density parameter, baryon fraction, the Hubble constant in units of 
$100km/s/\mathrm{Mpc}$, 
the scalar amplitude normalized at $k_0=0.002\Mpci$ and its power index, respectively. 
We also consider different cosmologies to check the generality of the result.
Since we can check the dependence of the response function on the overall amplitude of the power spectrum
by looking at the results at different redshifts, we here focus on the variety in only the shape of the spectrum.
As a representative of the parameters that control the shape, we consider the spectral tilt $\ns$. 
We run simulations for two additional models, one with a higher ($1.21$; \texttt{high\_ns}) and the other with
a lower ($0.71$; \texttt{low\_ns}) value of $\ns$.
Although the parameter $\ns$ has been constrained very tightly from observations of the cosmic 
microwave background (with only $\sim1\%$ uncertainty), we choose to give it a rather large ($\pm0.25$)
variation to cover a wider class of models with different linear power spectra.
The amplitude parameter $\As$ for these models are determined such that the rms linear fluctuation at 
$8\,h^{-1}\mathrm{Mpc}$ is kept unchanged.
The matter transfer function is computed for these models using the \texttt{CAMB} code \cite{CAMB}
with the high-precision mode of the calculation in the transfer function (\texttt{transfer\_high\_precision}
is set to be true and \texttt{accuracy\_boost}$=2$) up to 
$k_\mathrm{max} = 100\,h\,\mathrm{Mpc}^{-1}$. We confirm that the result is well converged by testing
more strict values in the parameter file.

We run four sets of simulations for the fiducial model with different volume and number of particles 
as listed in Table~\ref{tab:sim}. 
Covering different wave number intervals, these simulations allow us to examine the convergence of the 
measured response function.
The initial conditions are created using a code developed in \cite{Nishimichi09,Valageas11a}
based on the second-order Lagrangian PT (e.g., \cite{Scoccimarro98,Crocce06a}).
The initial redshifts of the simulations are determined as follows.
A lower starting redshift can induce transient effects associated with higher-order decaying modes.
On the other hand, as increasing the initial redshift, the randomly generated particle position 
generally gets closer to the pre-initial grid, and this can lead to discreteness noise in the force calculation.
To minimize the sum of these two systematic effects, we set the initial redshift such that 
the rms displacement is roughly $20\%$ of the inter-particle
spacing, and thus it depends on the resolution as shown in Table~\ref{tab:sim}.
We evolve the matter distribution using a Tree-PM code \texttt{Gadget2} \cite{Gadget2}. 
We finally measure the power spectrum by fast Fourier transform of the Cloud-in-Cell (CIC) 
density estimates on $1024^3$ mesh with the CIC kernel deconvolved in Fourier space.

For each set of simulations, we prepare multiple initial conditions with linear spectra perturbed 
by $\pm1\%$ over $q_{j}\leq q <q_{j+1}$. The amplitude of perturbation should be sufficiently small
such that the correction from the higher-order derivative ($\delta^2 \Pnl /\delta\Plin \delta \Plin$)
does not contaminate the result. We tested different amplitudes ($\pm 3\%$ and $\pm 5\%$), 
and confirmed that the result is almost unchanged.
We set the bin width as $\Delta \ln q = \ln(\sqrt{2})$ and each simulation set covers different wavenumber 
range corresponding to the box size and resolution limit.
The binning effect will be taken into account in the analytical calculations for fair comparison.
For the best resolution run, \texttt{L9-N10}, we study only five bins on small scales.
Further, we perform four realizations for \texttt{L9-N9} and \texttt{L9-N8} at each wave-mode bin
to estimate the statistical scatter. 
The same random phases are used for initial conditions with positive and negative perturbations
at each $q$ bins for each realization. Since the estimator~(\ref{eq:estimator}) takes the difference
of the two spectra, this helps us to reduce the statistical scatter on the response function significantly.

\begin{table*}[t]
   \centering
   \caption{Simulation parameters. Box size (box), softening scale (soft) and mass of the particles (mass) are 
   respectively given in unit of $\hiMpc$, $\hikpc$ and $10^{10}\himsun$. The number of $q$-bins
   is shown in the ``bins" column, for each of which we run two simulations with positive and negative
   perturbations in the linear spectrum. The ``runs" column shows the number of independent 
   initial random phases over which we repeat the same analysis. 
   The total number of simulations are shown in the ``total" column. 
   \label{tab:sim}}
   \begin{tabular}{c|cccccccc}
      name & box & particles & $z_\mathrm{start}$ & soft & mass & bins & runs & total \\
      \hline
      \texttt{L9-N10} & 512 & $1024^3$ & 63 & 25 & 0.97 & 5 & 1 &10\\
      \texttt{L9-N9} & 512 & $512^3$ & 31 & 50 & 7.74 &15 & 4 &120\\
      \texttt{L9-N8} & 512 & $256^3$ & 15 & 100 & 61.95 & 13 & 4 & 104\\
      \texttt{L10-N9} & 1024 & $512^3$ & 15 & 100 & 61.95 & 15 & 1 & 30\\
      \texttt{high\_ns} & 512 & $512^3$ & 31 & 50 & 7.74 & 5 & 4 & 40\\
      \texttt{low\_ns} & 512 & $512^3$ & 31 & 50 & 7.74 & 5 & 4 & 40
   \end{tabular}
\end{table*}

\section{Shape of the response function and comparison with PT} 
\begin{figure}[!ht]
   \centering
   \includegraphics[width=7cm]{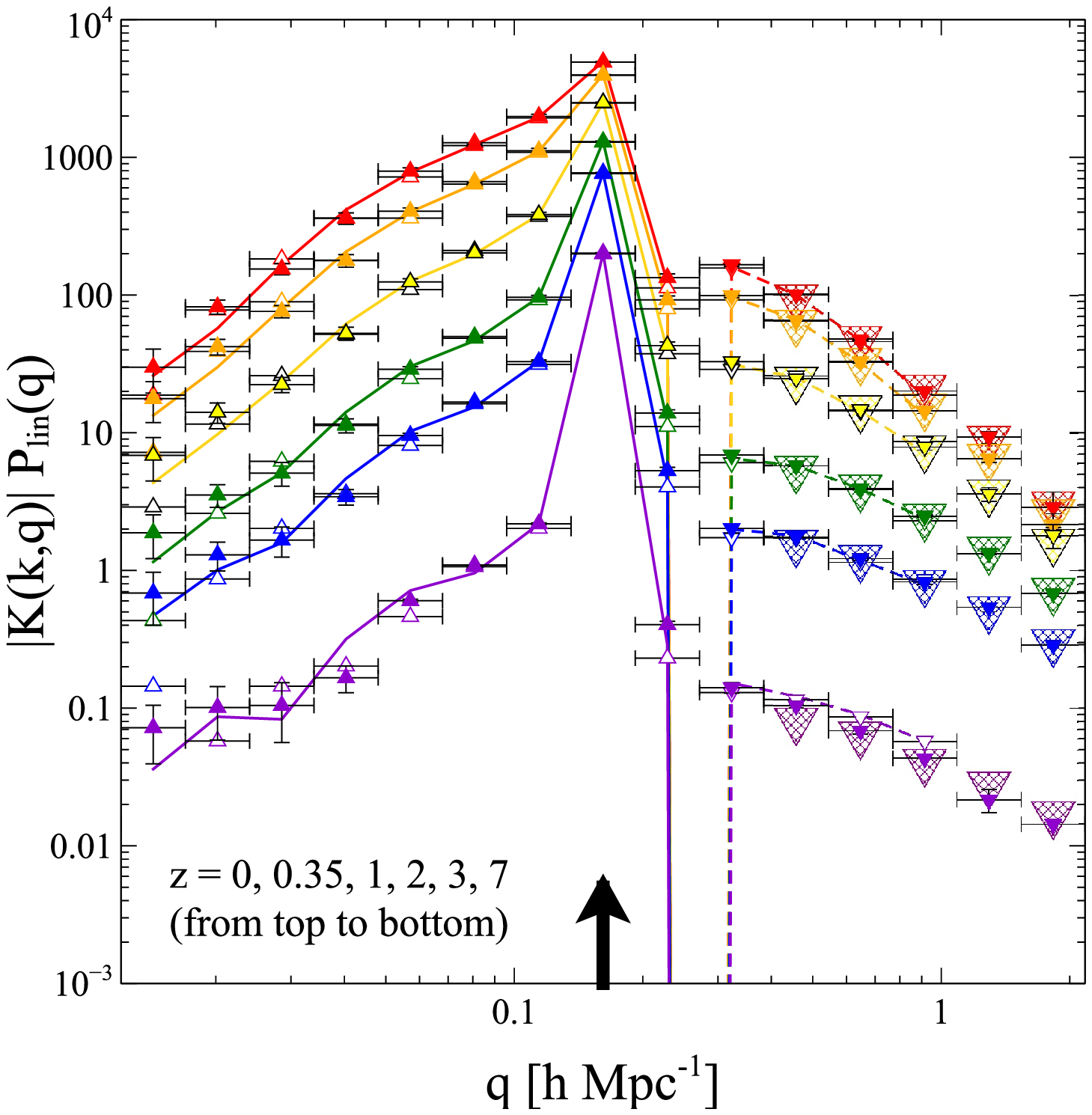}
   \caption{Response function measured from simulations. We plot $|K(k,q)|\Plin(q)$ as a function of the 
   linear mode $q$ for a fixed nonlinear mode at $k=0.161\,\hMpci$ indicated by the vertical arrow.
   The filled (open) symbols show {\tt L9-N9} ({\tt L10-N9}), the lines depict {\tt L9-N8},
   while the big hatched symbols on small scales are {\tt L9-N10}.
   Positive (negative) values are indicated as the upward (downward) triangles or the solid (dashed) lines. 
   }
   \label{fig:kernel}
\end{figure}

We are now in a position to present the response function measured from simulations.
The combination $K(k,q)\Plin(q)$ is plotted at a fixed $k$ shown by the vertical arrow 
as a function of $q$ in Fig.~\ref{fig:kernel}.
The strong overlap among different symbols and lines ensures the convergence of the results 
against resolution and volume. 

At high redshifts, we can see a prominent peak at $k=q$ as expected from linear theory (i.e., no mode transfer).
Nonlinear coupling then gradually grows with time and the peak feature gets less significant.
One of the key features here  
is the larger contribution from smaller wave modes ($q<k$); the growth
of structure is dominated by mode flows from large to small scales. 
Not surprisingly, the formation of a structure is more efficiently amplified when it is part of 
a larger structure than when it contains small-scale features.

\begin{figure}[!ht]
   \centering
   \includegraphics[width=7cm]{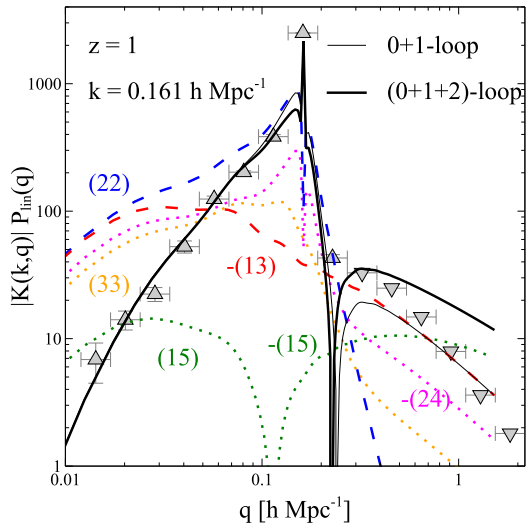}
   \caption{Response function predicted by PT (un-binned) up to one- (thin solid) 
   and two-loop (thick solid) order at $k=0.161 \hMpci$ at $z=1$.
   Dashed (dotted) lines show each of the one- (two-)loop contributions with the legend $(ij)$ showing the
   perturbative order of the calculation. We show a negative sign in the legend when $K$ is negative.
   The simulation data \texttt{L9-N9} are also shown by triangles.
}
   \label{fig:kernel_SPT}
\end{figure}

Such findings are fully in line with expectations from PT calculations. 
We show the predictions in Fig.~\ref{fig:kernel_SPT} up to the two-loop level 
(i.e., next-to-next-to-leading order) ignoring binning effects at this stage.
We present the contribution from $P_{ij}(k)\propto\langle\delta^{(i)}\delta^{(j)}\rangle$, 
where $\delta^{(i)}$ is the $i$th-order overdensity in the PT expansion.
The terms at the same loop order cancel at small $q$ due to the galilean invariance
of the system as discussed in 
e.g.,~\cite{Jain:1996lr,Scoccimarro:1996fk,Peloso:2013fk,Kehagias:2013lr,Blas:2013qy}.
On the other hand, small scales are dominated by one term at each order, 
$P_{13}(k)$ and $P_{15}(k)$. 
Similarly, it has been shown that, at the $p$-loop order in PT, the term originating from
the $(2p+1)$th-order density kernel function, $F_{2p+1}$,
dominates the mode-coupling effect from small scales~\cite{Bernardeau:2014lr}.

\begin{figure}[!ht]
   \centering
   \includegraphics[width=7cm]{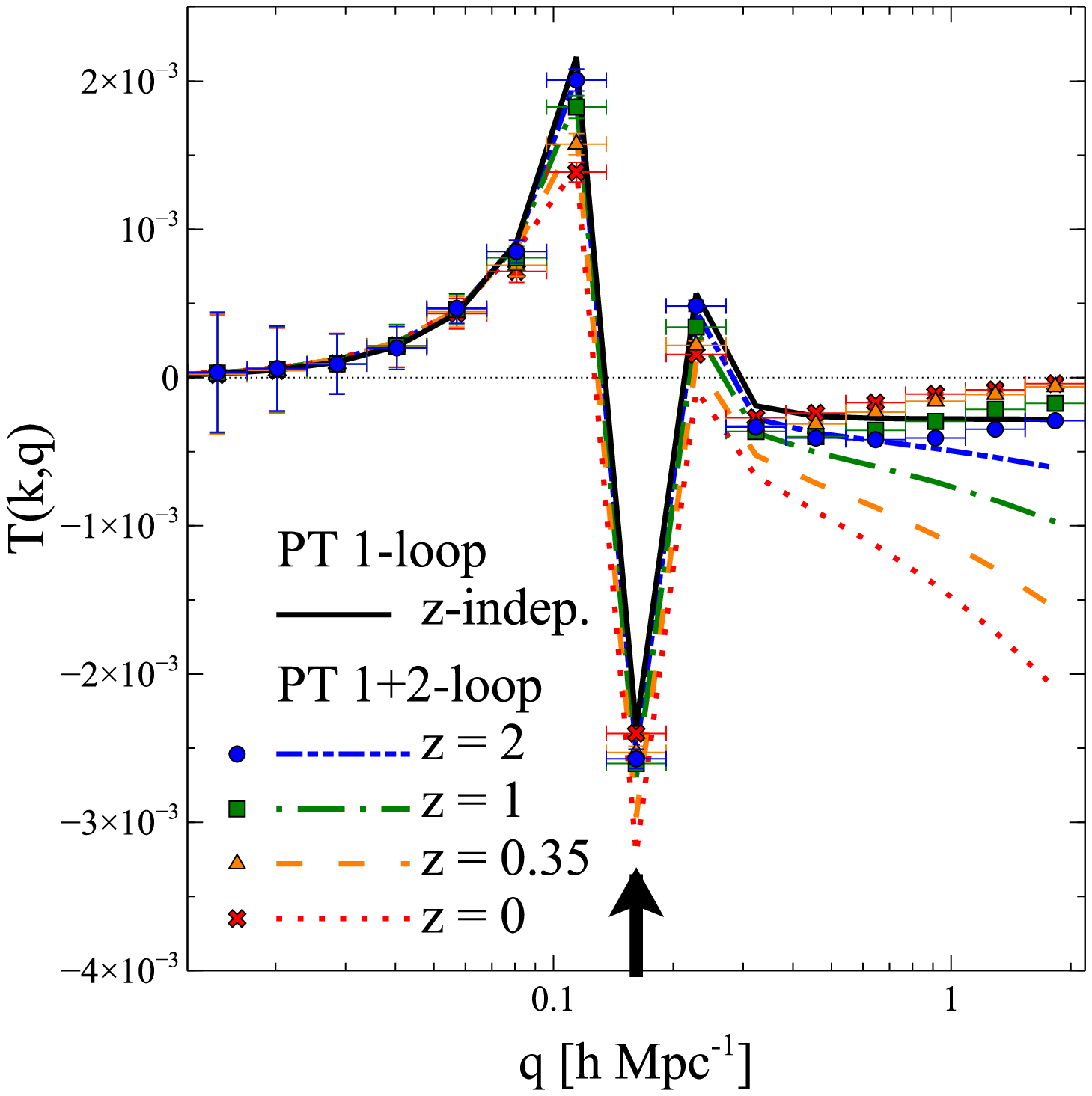}
   \caption{Rescaled response function, $T(k,q)\equiv [K(k,q)-K^\mathrm{lin}(k,q)]/[q\Plin(k)]$.
   PT calculations are shown by lines, whereas the symbols are 
   \texttt{L9-N9} (see legend for detail). The nonlinear wave-mode bin 
   is fixed at $k = 0.161\,\hMpci$ (vertical arrow).
   Binning is taken into account to the analytical calculations
   consistently to the simulations.
}
   \label{fig:T}
\end{figure}

We then rescale the response function at various redshifts as $T(k,q)=[K(k,q)-K^\mathrm{lin}(k,q)]/[q\Plin(k)]$, 
where $K^\mathrm{lin}$ is the linear contribution, and plot them in Fig.~\ref{fig:T}.
They are compared with the one-loop PT calculation (solid), 
which is time-independent with this normalization.
The simulation data indeed show little time dependence at $q\lesssim k$ in remarkable agreement with 
the one-loop calculation, reproducing the expected $q$ dependence~\footnote{It has the small-$q$ asymptote 
$[2519/4410-(23/84)n+(1/20)n^2]q^2/\pi^2$ for an Einstein-de Sitter background, 
with $n$ being the local slope of the linear spectrum.}, as well as the change of sign between large and small scales.
The small but non-negligible $z$-dependence at $k\sim q$ is further reproduced by the two-loop 
calculation (see the figure legend).
Note that at the wave-mode $k$ plotted here (i.e., $0.161\,\hMpci$), the two-loop PT prediction for
the nonlinear power spectrum agrees with simulations within $1\%$ at $z\gtrsim1$ and the agreement gets worse
at lower redshift reaching to $\sim5\%$ at $z=0$ (see e.g., \cite{Blas14}).

At $q\gtrsim0.3\,\hMpci$, however, the measured response function is damped compared to the PT.
The one-loop PT predicts the response function to 
reach a constant~\footnote{This constant is $-61k^2/(630\pi^2)$ for an Einstein-de Sitter background.}; 
at the two-loop order, it grows in amplitude with time. The numerical 
measurements show on the other hand that the scaled response function is strongly damped with decreasing redshift. 
It is such that the couplings take place effectively between modes of similar wavelengths. 
This effect is particularly important at late time. At redshift zero, 
the discrepancy between the model and simulations is striking. 
Furthermore, analysis of the response structure at three and higher loop order
(see e.g.,~\cite{Bernardeau:2014lr})
suggests that PT calculations, at any finite order, predict an even larger amplitude 
of the response function in the high $q$ region.  
This strongly suggests that this anomaly is genuinely non-perturbative.

\begin{figure}[!ht]
   \centering
   \includegraphics[width=7cm]{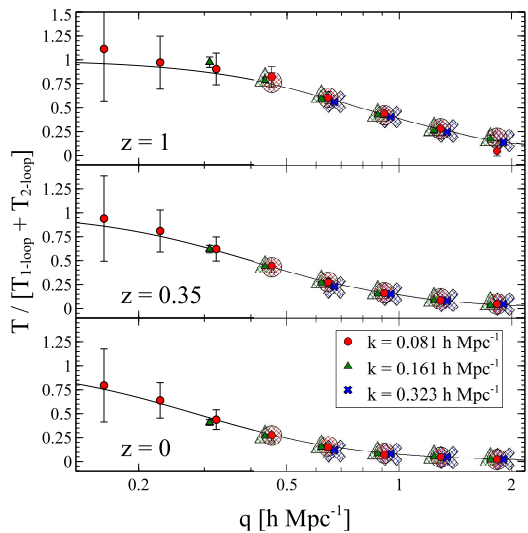}
   \caption{Response function divided by the two-loop PT at the three wave modes $k$ shown in the legend.
   We plot data points only at $q\geq 2k$ for definiteness. The over-plotted solid lines
   correspond to the empirical form~(\ref{eq:Teff}),
   small solid symbols are \texttt{L9-N9} while the big hatched are {\tt L9-N10}.}
   \label{fig:kernel_damp}
\end{figure}

\begin{figure}[!ht]
   \centering
   \includegraphics[width=7cm]{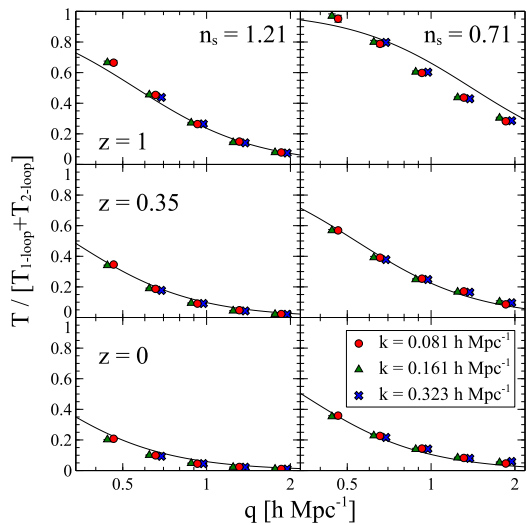}
   \caption{Same as Fig.~\ref{fig:kernel_damp}, but for cosmological models with different spectral indices,
   \texttt{high\_ns} and \texttt{low\_ns}.}
   \label{fig:kernel_damp_ns}
\end{figure}

We propose an effective description of this observed behavior. As illustrated in 
Fig.~\ref{fig:kernel_damp} it can be modeled with a Lorentzian: 
\begin{equation}
T^{\rm eff.}(k,q)\xrightarrow{\mathrm{high}-q}T^{\rm 1+2-loop}(k,q)\frac{1}{1+(q/q_{0})^{2}}
\label{eq:Teff}
\end{equation}
characterized by a critical wave mode, $q_{0}$, which does not depend on the nonlinear wave mode $k$.
We naively expect that this scale corresponds to the scale at which the fluctuation is order unity and
thus pertrubative expansion is not valid.
Indeed, we numerically found that a simple fitting formula
\begin{equation}
\sigma_{\mathrm{lin}}(R;z)|_{R=1/q_0} = 1.35,
\label{eq:q0}
\end{equation}
where $\sigma_{\mathrm{lin}}^2(R)$ is the variance of the linear density fluctuation
smoothed with a gaussian filter of the form $\exp[-(0.46kR)^2/2]$
\footnote{The factor $0.46$ here is chosen for a better correspondence with the top-hat radius in terms of the variance. See \cite{Paranjape:2012aa} 
for more detail.},
can explain reasonablly well the data points not only
for the fiducial model but for the models with different spectral indices (see Fig.~\ref{fig:kernel_damp_ns}).
One can find that the fit is not as accurate at $z=1$ for the \texttt{low\_ns} model, suggesting the limitation of the fitting formula.
Nevertheless, a simple form (\ref{eq:Teff}) with a single 
parameter $1.35$ in Eq.~(\ref{eq:q0}) seems to capture the damping tail 
of the response function in a rather wide range of cosmological models at
different redshifts.
Note that, the $k$-dependence of the responce function at the high-$q$ limit 
is preserved in perturbative calculations (it is always proportional to $k^2$ independent 
of the perturbative order due to the asymptote of the $F_n$ kernel function).
The independence of $q_0$ on $k$ is thus in full agreement with PT predictions.

\section{Discussion}
The simulation results give a clear evidence that the mode transfer from small 
to large scales is suppressed compared to the PT prediction when the mode $q$ enters the 
nonperturbative regime. However, the origin of the suppression is yet to be understood.
In particular it is not clear whether this has its roots in shell 
crossing or not. For instance, we can find in \cite{Pueblas:2008aa} that vorticity and velocity 
dispersion generated by shell crossing can alter the evolution of density fluctuations through 
nonlinear coupling.
Especially, the latter is shown to give non-negligible corrections to the density power spectrum
even on very large ($\sim 0.1\hMpci$) scales. 
This indicates that multistreaming physics, which take place on small scales, are somehow related to the growth of 
large-scale fluctuations, and this is exactly the coupling of large and small scales that we discuss here.
Effective Field Theory (EFT) approaches as advocated originally 
in \cite{Baumann12,Carrasco12,Hertzberg:2014lr} would be a natural framework to invoke 
for accounting for such multistreaming effects.
In these approaches, however, the response function is ultimately encoded in free coefficients for which 
no theory exists.

It might be possible that such damping effect 
originates from simpler mechanisms in single-stream physics. 
It has been shown in particular that the nonlinear density propagator, which expresses the evolution of a given wave mode 
with time, is exponentially damped by the large-scale displacements. This is the standard result
on which the Renormalized Perturbation Theory is based \cite{Crocce06b,Crocce:2006uq}. 
As explicitly shown in \cite{Bernardeau12} equal-time spectra are however insensitive to displacements of 
the global system, that originates from wave modes smaller than $k$. 
Displacements at intermediate scales are nonetheless expected to induce some effective 
damping for equal-time spectra. The physical idea behind that is that the force driving the collapse of a large-scale
perturbation (e.g., a cluster of galaxies) is affected by the small scale inhomogeneities within the structure (say galaxies), 
but that this dependence might be damped when such small scale inhomogeneities are actually moving within the structure. 
It is however beyond the scope of this presentation to evaluate the importance of this effect.

\section{Summary}
We have presented the first direct measurement of the response function 
that governs the dependence of the nonlinear power spectrum on the initial spectrum
during cosmic structure formation.
This measurement was done 
using a large ensemble of $N$-body simulations that differ slightly in their initial conditions.
The results were found to be robust to the simulation resolution -- 
as shown in Table \ref{tab:sim} -- supporting 
the idea that measured shapes were genuine features in the development of gravitational instabilities.

The response functions were computed concurrently at next and next-to-next leading order in PT. Comparisons
with measurements show a remarkable agreement over a wide range of scale and time. 
We found however mode transfers from small to large scales to be strongly suppressed compared to theoretical 
expectations especially at late time. We propose a description of the damping tail with a Lorentzian shape.

These results are of far-reaching consequences. They first give insights into the mode coupling structure of
cosmological fluids and show that PT approaches capture most of their properties. The small scale damping 
signals the validity limit of the PT beyond next-to-leading order. 
It provides in particular indications on how to regularize their contributions.
The observed damping also marks the irruption of collective non-linear effects although the underlying mechanisms are yet to be uncovered.
Most importantly the damped response suggests
that small scale physics, whether from the initial metric perturbations or late-time processes, can be effectively controlled. It
paves the way for solid estimates of the theoretical uncertainties on the determination of
cosmological parameters (such as inflationary primordial non-Gaussianities, neutrino masses or dark energy parameters) 
from large-scale surveys.

\section*{Acknowledgment}
We thank Patrick Valageas for fruitful discussions on analytical calculations of the response function.
We also thank Yasushi Suto for insightful comments on the origin of our findings.
This works is supported in part by grant ANR-12-BS05-0002 of the French Agence Nationale de la Recherche.
TN is supported by JSPS.
AT is supported by a Grant-in-Aid for Scientific Research from the Japan Society for the Promotion of 
Science (JSPS) (No.~24540257).
Numerical calculations for the present work have been carried out on Cray XC30 at 
Center for Computational Astrophysics, CfCA, of National Astronomical Observatory of Japan.


\bibliography{ms.bbl}

\end{document}